\documentclass[referee]{aa_modified} 
\usepackage[colorlinks=true,linkcolor=blue, citecolor=blue, filecolor=blue, urlcolor=blue]{hyperref}
\usepackage{graphicx}
\usepackage{txfonts}
\begin{document}
   \title{Preferential acceleration of heavy ions in magnetic reconnection: 
   Hybrid-kinetic simulations with electron inertia}


   \author{Neeraj Jain
          \inst{1}
          \and
          J\"org B\"uchner\inst{1}
          \and
          Miroslav B\'arta\inst{2}
          \and
          Radoslav Bu\u{c}\'ik \inst{3}
          }

   \institute{Zentrum f\"ur Astronomie und Astrophysik, Technische 
   Universit\"at Berlin, Hardenbergstr. 36, D-10623, Berlin, Germany\\
              \email{neeraj.jain@tu-berlin.de}
         \and
             Astronomical Institute of the Academy of Sciences of the Czech 
             Republic, Fri\u{c}ova 298, Ond\u{r}ejov, 251 65, Czech Republic\\
          \and
          Southwest Research Institute, 6220 Culebra Road, San Antonio, TX 
          78238, USA
             }

   \date{\today}

 
  \abstract
   {Solar energetic particles (SEPs) in the energy range 10s KeV/nucleon - 100s 
   MeV/nucleon originate  from Sun.  Their high flux near Earth may damage the 
   space borne electronics and generate secondary radiations harmful for the 
   life on Earth and thus understanding their energization on Sun is important 
   for space weather prediction. Impulsive (or ${}^{3}$He-rich) SEP events are 
   associated  with the acceleration of charge particles in solar flares by 
   magnetic reconnection and related processes. The preferential acceleration 
   of heavy ions  and the extra-ordinary abundance enhancement of ${}^3$He in 
   the impulsive SEP events are not understood yet.}
   {In this paper we study acceleration of heavy ions and its consequences for 
   their abundance enhancements by magnetic reconnection, an established 
   acceleration source for
   	impulsive SEP events in which heavy-ion enhancement is observed
   	}
   {  We employ a two 
   	dimensional hybrid-kinetic plasma model (kinetic ions and inertial 
   	electron fluid) to simulate magnetic reconnection. All the ions species are 
   	treated self-consistently in our 
   	simulations.}
   {We find that heavy ions are preferentially accelerated to energies many 
   times larger than their initial thermal energies by a variety of 
   acceleration mechanisms operating in reconnection.  Most efficient 
   acceleration takes place in the flux pileup regions of magnetic 
   reconnection. Heavy ions with sufficiently small values of charge to mass 
   ratio   ($Q/M$) can be accelerated by pickup mechanism in outflow regions 
   even before any magnetic flux is piled up. The energy spectra of heavy ions 
   develop  a shoulder like region, a non-thermal feature, as a result of the 
   acceleration. The spectral index of the power law fit to the shoulder region 
   of the spectra varies approximately as $(Q/M)^{-0.64}$. Abundance 
   enhancement factor, defined as number of particles above a threshold energy 
   normalized to total number of particles, scales as $(Q/M)^{-\alpha}$ where 
   $\alpha$ increases with the energy threshold. We discuss our simulation 
   results in the light of the SEP observations.}
   {}

\titlerunning{Preferential acceleration of heavy-ions in reconnection}
\maketitle   
%

\section{Introduction}

Non-thermal acceleration of charged particles is a widespread process in space 
and astrophysical plasmas ranging from planetary magnetospheres to clusters of 
galaxies. The accelerated particles, mostly ions, are detected near and on 
Earth in a wide energy range , $10^4-10^{20}$ eV,  either directly by particle 
detectors on satellites and high altitude balloons or indirectly by detecting 
the secondary particles and electromagnetic radiation produced by their 
interaction with Earth's atmosphere. These particles are generally categorized 
as solar energetic particles (SEPs), galactic cosmic rays and extra-galactic 
cosmic rays  based on their source regions in our star the Sun, our galaxy 
Milky Way  and beyond our galaxy, respectively. Solar energetic particles 
(SEPs) in typical energy range of 10s KeV/nucleon - 100s MeV/nucleon are of 
particular interest because their flux near Earth is sufficiently high to have 
implications for space weather phenomena.

SEP events can be classified into impulsive (short duration $\leq$ 1 day, less 
intense --- relatively smaller particle fluxes with typical energies $< 10$ 
MeV/nucleon and numerous --- about 1000 per year) and gradual (long duration 
$\sim$ several days, orders of magnitude more intense --- large particle fluxes 
with energies $>$ 10 MeV/nucleon, less frequent $\sim$ 10 per year  ) events 
\citep{Reames1999,Reames2021_book}. 
 Gradual events are attributed to the acceleration of charged particles by 
	Coronal Mass Ejection (CME) driven shocks while impulsive events to the 
	acceleration in solar flares by magnetic reconnection and associated 
	processes 
	\citep{kahler2001,Mason2007,Drake2009,lin2011,Reames2013,murphy2016,Desai2016_review,Bucik2020,reames2022}.
Impulsive events show abundance 
enhancements of 
heavier ions relative to their abundances in the solar corona  at energies well 
above their average thermal energy ($\sim$ 100 eV) \citep{Reames2021_review}. 
	In contrast to impulsive events, average abundances of elements in gradual 
	events are similar to solar corona, e.g., \citep{reames2018}.  
The enhancement factor, defined as the ratio of the relative (to a reference 
element, usually Oxygen) abundances of an element X in impulsive SEP events and 
solar corona, exhibits power law dependence on ion's charge ($Q$) to mass ($M$) 
ratio as $(Q/M)^{-\alpha}$ for $M \ge 4$ amu. Event-to-event variations 
in the value of $\alpha$ 
have been found \citep {Reames2014a} with mean values $\alpha=3.26$ (at 0.375 
MeV/nucleon) \citep{Mason2004}, $\alpha=3.64 \pm 0.15$ (at 3-10 MeV/nucleon) 
\citep{Reames2014}, and $\alpha=3.53$ (at 160-226 KeV/nucleon) and 
$\alpha=3.31$ (at 320-453 KeV/nucleon) \citep{Bucik2021}. 
Enhancement factor for ${}^3$He isotope, however, does not obey the power law 
and can have very large values (up to $10^4$) compared to those calculated 
using the power law  \citep{Kocharov1984,Mason2007}. For this reason, impulsive 
SEP events are also called ${}^3$He-rich events. The enhancement of heavy ions 
is, however, not correlated with the extra-ordinary enhancement of ${}^3$He 
\citep{Mason1986,Reames1999}. 
The abundance enhancements of heavy ions with small values of $Q/M$ and the extra-ordinary abundance 
enhancement of ${}^3$He in impulsive SEP events are not understood yet.
Since these enhancements are un-correlated \citep{Mason1986,Reames1999}, 
attempts have been made to explain the abundance enhancement of heavy ions 
and ${}^3$He by separate mechanisms. Abundance enhancement of heavy ions is  
considered to be due to  their preferential acceleration by magnetic 
reconnection and the associated processes in solar flares  from 
thermal to SEP energies.
Models of the preferential acceleration of heavy ions consider  resonant 
interaction of the ions with the waves in Alfv\'enic plasma turbulence \citep{Miller1998,Eichler2014,Kumar2017,Fu2020,Shi2022}  which may be generated during magnetic reconnection, e.g., development of turbulent outflows  as shown by fully kinetic 3-D simulations of magnetoc reconnection \citep{lapenta2020}.  
	Turbulent energy 
decays with wave frequency, i.e., higher frequency waves have lower energy. 
Heavy ions with lower values of $Q/M$ have lower values of cyclotron frequency 
($\propto Q/M$) and thus would resonate with lower frequency but higher power 
waves in turbulence, favoring preferential acceleration of heavy ions. 
 In addition, turbulence amplitude above a threshold can make ions' orbit chaotic heating ions stochastically. Phenomenological arguments and simulations of test particles interacting with turbulent spectrum of Alfv\'en and kinetic Alfv\'en waves at wavelengths comparable to ion gyro-radius in plasmas with $\beta < 1$ found that, at comparable temperatures, alpha particles and heavier ions are stochastically heated more efficiently than the protons \citep{chandran2010}. 
	Three dimensional hybrid-kinetic simulations of continuously driven Alfv\'enic turbulence at low plasma beta show that the Hall and thermoelectric effects in a generalized Ohm’s law contribute significantly to the stochastic ion heating which is enhanced by the intermittency in the turbulence \citep{cerri2021a}. 
%
%
The 
mechanisms proposed for the preferential acceleration of ${}^3$He consider 
absorption of some wave energy by cyclotron resonance of  ${}^3$He with the 
wave \citep{Fisk1978, Temerin1992, Liu2006}. These waves can be produced by 
electron beams, e.g., by electrons streaming along open magnetic field lines in 
solar corona, or via coupling with low-frequency Alfv\'en waves. In fact, 
efficient acceleration of ${}^3$He by ion-cyclotron resonance has been observed 
in nuclear fusion devices, which has implications for space plasmas as well 
\citep{Kazakov2017}.

 Magnetic reconnection can also accelerate electrons and ions  in solar 
	flares 
	by non-resonant mechanisms 
	which have also been 
	considered to explain the preferential acceleration of heavy ions 
	\citep{Drake2009,Barta2011a,Barta2011b,Zhou2015,Zhou2016}.
In 2.5-D magnetohydrodynamic (MHD) and test particle simulations 
\citep{Kramolis2022}, heavy ions were found to be preferentially accelerated  
by first order Fermi process in cascading plasmoids  generated by spontaneous 
magnetic reconnection in a meso-scale current sheet. The ion energy spectra and 
abundance enhancement factors exhibit power-law profiles. The index of the 
power law for the abundance enhancement factor, however, was not in agreement 
with the observations. The authors suspected that the disagreement could be due 
to the limitation of the MHD model which lacks the kinetic physics essential 
for magnetic reconnection in collisionless plasmas.  
In 2-D fully kinetic simulations of magnetic reconnection, ions entering the 
reconnection exhaust can behave like pickup particles and gets accelerated if 
their $Q/M$ is below a threshold value, and thus preferential acceleration of 
heavy ions. An explanation for the power-law dependence of the enhancements on 
$Q/M$ was proposed based on the pickup mechanism 
\citep{Drake2009,Knizhnik2011}. These simulations, carried out for only for one 
ion species (${}^4$He$^{2+}$), did not verify the power-law dependence. It is, 
therefore, not clear if acceleration of heavy ions by pickup mechanism can 
provide the observed scaling of the abundance enhancement with $Q/M$.  
Magnetic reconnection in turbulence generated current sheets, which form in turbulence as a result of the magnetic energy cascade from large to kinetic scales where magnetic fluctuations have intermittent statistics  \citep{cerri2019}, have also been suggested to  contribute to energization of charged particles \citep{rueda2022,groselj2017}. 

Magnetic reconnection in collisionless plasmas is a multi-scale process which 
occurs at electron kinetic scales and then couples to ion and even larger 
macro-scales. An ideal simulation of magnetic reconnection requires kinetic 
treatment of electrons and ions and covering scales from electron kinetic 
scales all the way up to large macro-scales well above the ion kinetic scales. 
Such fully kinetic simulations, e.g., using Particle-in-Cell (PIC) method, of 
magnetic reconnection in electron-proton plasma are computationally very 
demanding. Self-consistent inclusion of heavier ion species make the 
simulations even more demanding because now one has to use larger simulation 
box to accommodate the larger gyro-radii of the heavier species while at the 
same time resolve the electron scales. 
For this reason, a variety of reduced kinetic models have been used for numerical studies of magnetic reconnection, plasma turbulence and the associated charge particle heating. 
 Plasma turbulence at ion scale was studied using  hybrid-kinetic (with and without electron inertia) along with fully kinetic simulations \citep{cerri2019}.	
	Two dimensional fully kinetic, gyro-kinetic and hybrid-kinetic simulations of plasma turbulence  suggest that magnetic reconnection can heat electrons and ions in  turbulence generated current structures \citep{groselj2017}. 
Three dimensional hybrid-kinetic simulations of driven Alfv\'en-wave turbulence show a flattened core in the perpendicular-velocity distribution and non-Maxwellian wings in the parallel-velocity distribution of ions resulting from the dissipation of kinetic Alfv\'en waves at sub-ion scales \citep{arzamasskiy2019}.  
Two dimensional  simulations of guide field magnetic reconnection by means of  a  hybrid-Vlasov-Landau-fluid (HVLF) model  reproduces the main features of magnetic reconnection and anisotropic heating of electrons \citep{finelli2021}. 
In hybrid-kinetic simulations (with massless electron fluid) of plasma turbulence,  preferential heating of alpha particles with respect to protons was  observed near current sheets in spatial regions of enhanced ion vorticity \citep{valentini2016}.

  In this paper, we employ 2-D hybrid-kinetic simulations (kinetic ions and inertial electron fluid) to study acceleration of heavy ions and its consequences for 
	their abundance enhancements by magnetic reconnection, an established 
	acceleration source for impulsive SEP events in which heavy-ion enhancement 
	is observed 
	\citep{kahler2001,Mason2007,Drake2009,lin2011,murphy2016,reames2022}. Treatment of electrons as an inertial fluid allows to include the 
	physics at electron inertial scale but relaxes the numerical requirement of 
	resolving the Debye length in PIC method and therefore allowing larger 
	simulation domains for the same number of grid points.  The computational  feasibility, however, comes at the cost of electron kinetic physics which 
	is acceptable as far as the study of the ion acceleration is the primary 
	objective. 
All the ion species are treated self-consistently in our simulations. We use 
the term ``heavy ion'' to mean any ion species whose mass $M$ is larger than 
the proton's mass. The paper is organized as follows. Section 
\ref{sec:simulation_setup} presents the simulation setup.  Simulation results 
are presented in section \ref{sec:results} and discussed and concluded in 
section \ref{sec:conclusion}.

\section{Simulation setup\label{sec:simulation_setup}}
We carry out 2-D hybrid-kinetic simulations of magnetic reconnection with 
electron inertia using the hybrid-PIC code CHIEF (Code Hybrid with Inertial 
Electron Fluid) which is a 3-D code parallelized based on Message Passing 
Interface (MPI) for high performance computing 
\citep{Munoz2018,Jain2022,Munoz2023}. In the hybrid code CHIEF, electrons are 
treated as an inertial and isothermal fluid whose equations are coupled to 
Maxwell's equation to obtain the electric and magnetic fields. All the ions 
(protons and heavier ions) , on the other hand, are treated self-consistently 
as kinetic species  and their equations of motion in electric and magnetic 
fields are solved using the PIC method. The code CHIEF treats the inertial 
effects of the electron fluid without any of the approximations used by other 
electron-inertial hybrid-kinetic code \citep{Jain2022}.  The details of the 
hybrid-kinetic model used in CHIEF, its numerical implementation and 
parallelization are discussed in our other publications 
\citep{Munoz2018,Jain2022,Munoz2023}.

We initialize the 2-D simulations with two Harris current sheets, 
$\mathbf{B}=\hat{z}B_{z0}[\tanh\{(y+L_y/4)/L\}-\tanh\{(y-L_y/4)/L\}-1]+\hat{x}B_{x0}$,
 in a y-z plane and a guide magnetic field $B_{x0}=0.2\,B_{z0}$ perpendicular 
to the plane. The half-thickness $L=d_p$ of the current sheets is taken to be 
equal to a proton inertial length ($d_p$) and $L_y$ is the length of the 
simulation domain along y-direction. A small initial perturbation is added to 
the Harris equilibrium to form an X-point and an O-point in the current sheets 
centered at $y=-L_y/4$ and $y=L_y/4$, respectively. The plasma of Harris 
current sheets consists of quasi-neutral populations of proton particles and 
electron fluid with Harris equilibrium density  profile 
$n_e=n_p=n_0[\text{sech}^2\{(y+L_y/4)/L\}+\text{sech}^2\{(y-L_y/4)/L\}]$. 
Harris sheets are embedded in a uniform background plasma of density $0.2\,n_0$ 
and therefore peak density at their centers is $1.2\,n_0$. The background 
plasma consists of electron fluid and particle populations of  heavy ions (only 
one species) and protons with densities $n_{be}=0.2\,n_0$, $n_{bi}=0.01\,n_0$ 
(5\% of the background plasma density) and $n_{bp}=n_{be}-Z\,n_{bi}$, 
respectively,  where $Z$ is the charge state of heavy ions with charge of heavy 
ion species given by $Q=Z\,e$.
The initial velocity distribution of the protons and heavy ions is Maxwellian. 
The initial  temperatures of electrons, protons and heavy ions are the same, 
viz., $T_e=T_p=T_i=0.25\, m_p V_{Ap}^2$ where $m_p$ is proton mass  and 
$V_{Ap}=B_{z0}/\sqrt{\mu_0n_0m_p}$ is the proton Alfv\'en velocity based on 
$B_{z0}$ and $n_0$.

Each of our simulations consists of only one species of heavy ion. We consider 
the following four different species of heavy ions to be included (one at a 
time) in our simulations : ${}^4$He$^{2+}$, ${}^3$He$^{2+}$, ${}^{16}$O$^{7+}$ 
and ${}^{56}$Fe$^{14+}$.
We take proton to electron mass ratio as $m_p/m_e=25$. The simulation domain 
$L_y\times L_z$ is $51.2\,d_p \times 102.4\,d_p$ resolved by a grid spacing of 
$0.1\, d_p$ in each direction. The time step is $\Delta 
t=0.0025\,\omega_{cp}^{-1}$, where $\omega_{cp}=eB_{z0}/m_p$ is the proton 
cyclotron frequency. We take 500 and 200 particles per cell for protons and 
heavy ions, respectively. Plasma resistivity is taken to be zero and electron 
inertia allows the magnetic reconnection. Boundary conditions are periodic.


\section{Simulation results \label{sec:results}}
\begin{figure*}
	\includegraphics[clip=true,trim=0cm 0cm 0cm 0cm, 
	width=0.48\textwidth]{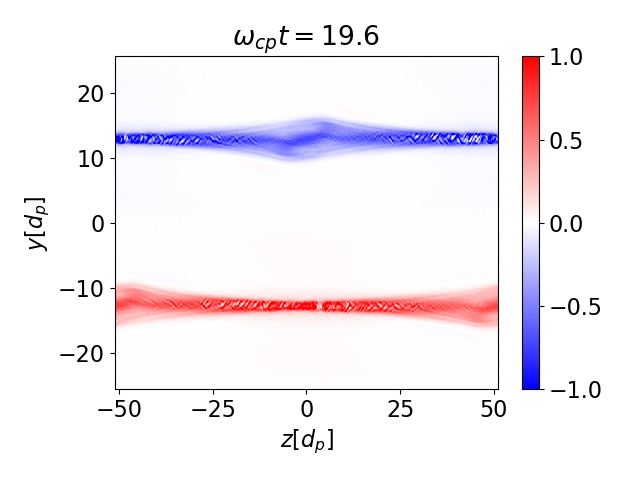}
	%
	\includegraphics[clip=true,trim=0cm 0cm 0cm 
	0cm,width=0.48\textwidth]{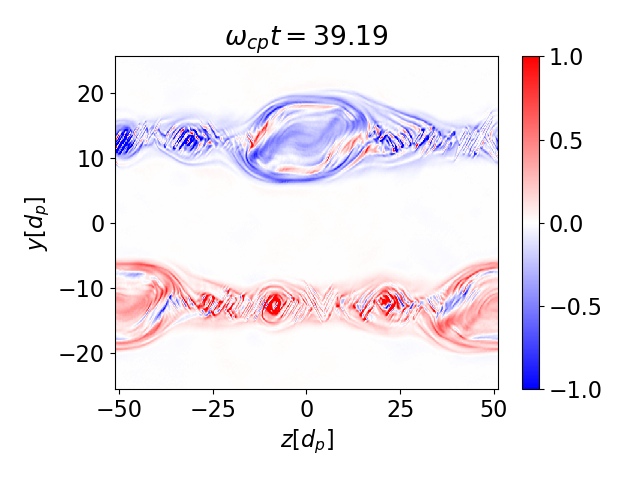}
	%
	\includegraphics[clip=true,trim=0cm 0cm 0cm 
	0cm,width=0.48\textwidth]{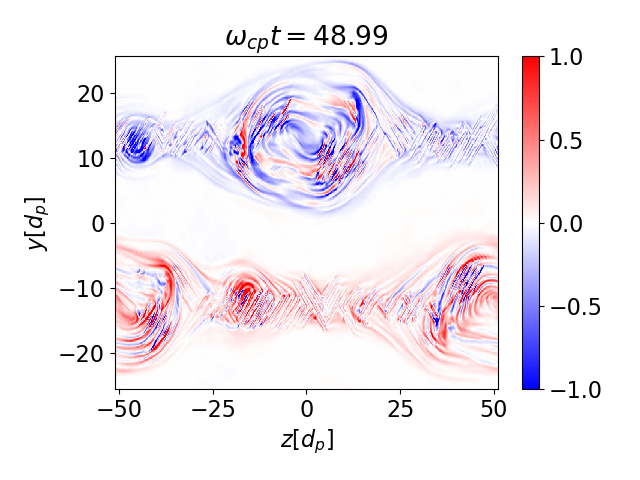}
	%
	\includegraphics[clip=true,trim=0cm 0cm 0cm 
	0cm,width=0.48\textwidth]{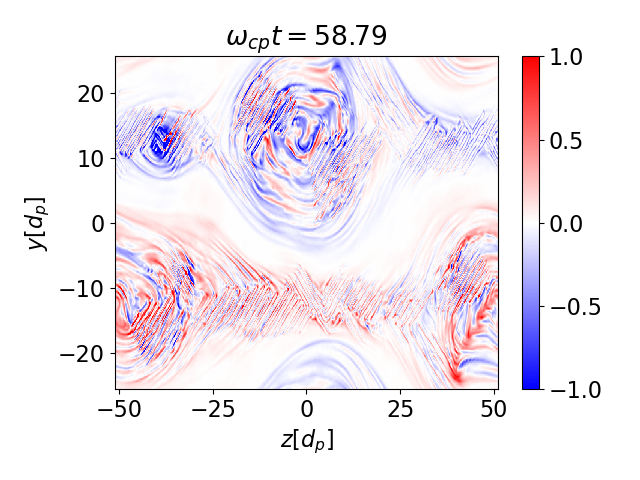}
	%
	\caption{Out-of-plane current density $J_x/(n_0eV_{Ap})$ at four different 
	times.\label{fig:jx_evolution}}
\end{figure*}

Fig. \ref{fig:jx_evolution} shows evolution of out-of-plane current density 
$J_x$ from initial to late phase. In the initial phase ($\omega_{cp}t=19.6$), 
an X-point forms in the lower current sheet ($y<0$) while an O-point forms in 
the upper current sheet ($y>0$), as per the  initialized perturbation. By 
$\omega_{cp}t=39.19$, the lower (also the upper) current sheet spontaneously 
develop magnetic islands as a result of magnetic reconnection at multiple 
sites, even though the initialized perturbation was chosen to initiate 
reconnection at a single site in the simulation domain. The magnetic islands on 
each of the current sheet grow in size with time by merging and/or pushing 
among themselves and simultaneously develop turbulence  inside them.
At $\omega_{cp}t=58.79$, the magnetic islands of the lower and upper current 
sheet grow to big enough size that the particles accelerated in the upper 
(lower) current sheet may cross over to the lower (upper) current sheet. 

Our objective here is to study the ion acceleration without the influence of 
the periodic boundaries. Since the upper current sheet is affected by the 
periodic boundary conditions along z-direction from the beginning (the initial  
X-points form at the z-boundaries), we focus on the lower current sheet for our 
studies. The lower current sheet is also likely to be affected by periodic 
boundaries along z-direction after a time $\sim 50 \omega_{cp}^{-1}$, Alfv\'en 
waves take to cross half the simulation domain.
In order to avoid particles crossing from the upper half to the lower half of 
the simulation domain and the influence of periodic boundaries,  we limit our 
analysis of results only up to the time $\omega_{cp}t=48.99$ and in the spatial 
region $-23 \leq y/d_p \leq -3$.



\begin{figure*}
	\centering
	\setlength{\unitlength}{0.1\textwidth}
	\begin{picture}(10,4)
	\put(0,0){\includegraphics[width=0.48\textwidth]{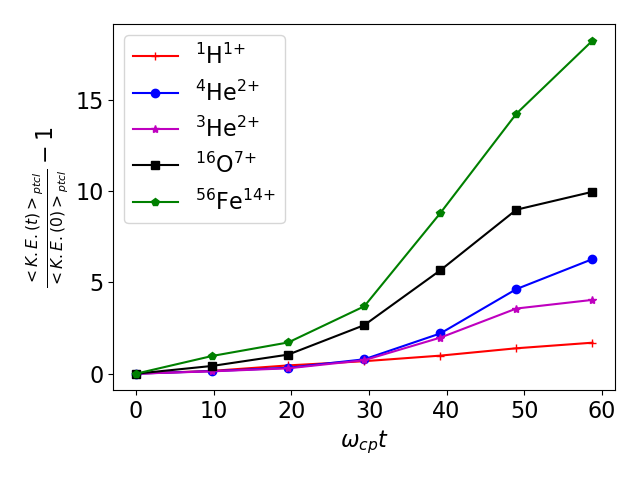}}
	\put(3,3){{\large (a)}}
	%
	\put(5,0){\includegraphics[width=0.48\textwidth]{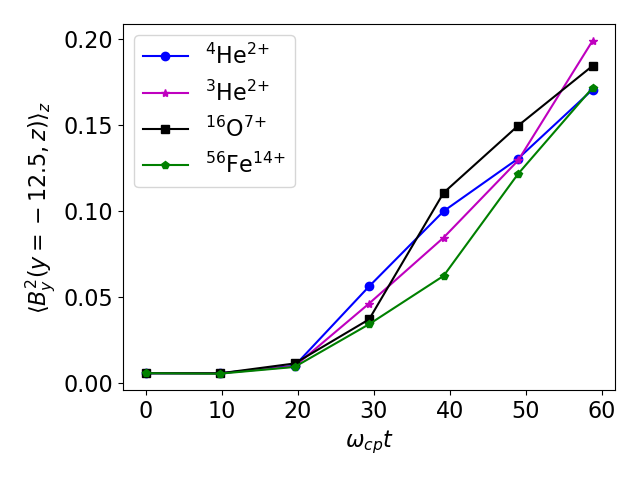}}
	\put(8,3){{\large (b)}}
	\end{picture}
	\caption{Time development of the fractional change in average kinetic 
	energy  per particle, $\langle K.E. (t) \rangle_{ptcl}$, from its initial 
	value (a) and average of the square of the normal component of the magnetic 
	field $B_y$ along the central line of the lower current sheet (b) for 
	different ion species. \label{fig:av_ke_evolution}}
\end{figure*}

Figure \ref{fig:av_ke_evolution}a shows the evolution of fractional change in 
average kinetic energy from its initial value
for different ion species. All the heavy ions gain energy first at slow rates  
up to $\omega_{cp}t\approx 30$ after which they gain energy at much faster 
rates.    Note that magnetic islands have significantly developed by 
$\omega_{cp}t=30$. This can be seen in Fig. \ref{fig:av_ke_evolution}b which 
shows that $\langle B_y^2 (y=-12.5\,d_p, z)\rangle_z/B_{z0}^2$ (average of the 
magnetic energy in the normal component of the magnetic field along the central 
line of the lower current sheet --- a proxy for the magnetic island 
development) begins to grow at $\omega_{cp}t=20$ and, by $\omega_{cp}t=30$, has 
grown to $\sim$ 5\% of the asymptotic magnetic energy in the anti-parallel 
magnetic field. The development of magnetic islands, therefore, seem to be 
linked with the efficient energization of heavy ions. Proton energization, on 
the other hand, does not seem to be affected significantly by the formation of 
magnetic islands as it does not enhance significantly after 
$\omega_{cp}t\approx 30$.

\begin{figure*}
	\centering
	\setlength{\unitlength}{0.05\textwidth}
	\begin{picture}(20,8)
	\put(0,0){\includegraphics[clip=true,trim=0.1cm 0cm 0cm 
	0cm,width=0.48\textwidth]{{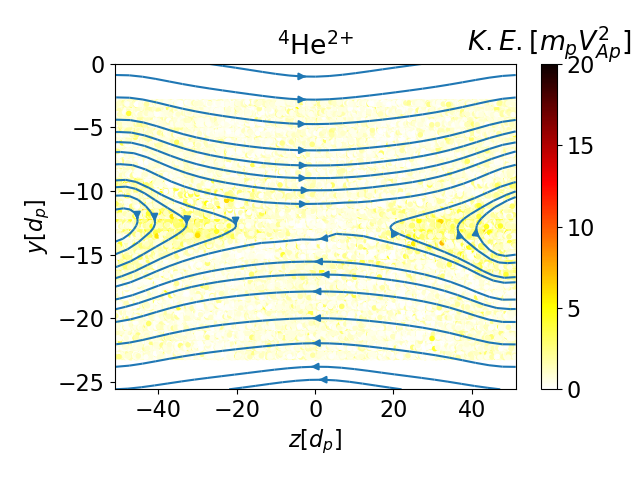}}}
	\put(4,7){{\large $\omega_{cp}t=19.6$}}
	\put(10,0){\includegraphics[clip=true,trim=0.1cm 0cm 0cm 
	0cm,width=0.48\textwidth]{{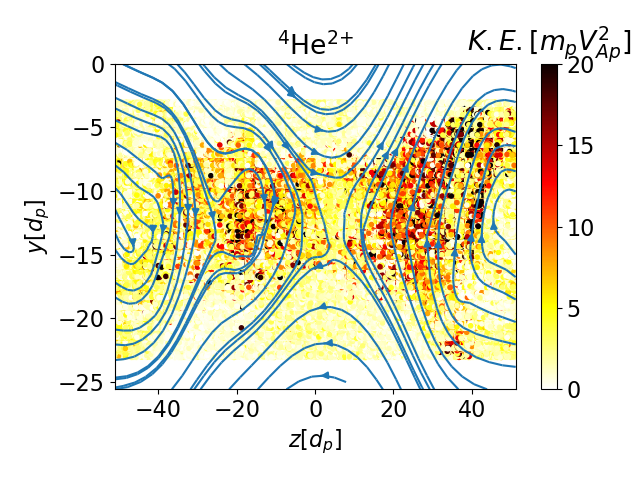}}}
	\put(14,7){{\large $\omega_{cp}t=48.99$}}
	\end{picture}
	\caption{Positions  of ${}^4$He$^{2+}$ ions represented by dots colored by 
		ion's kinetic energy at $\omega_{cp}t=19.6$ (left column) and 
		$\omega_{cp}t=48.99$ (right column).  Lines with arrows represent magnetic 
		field lines. Only the lower half ($y < 0$) of the simulation domain is 
		shown. \label{fig:acceleration_sites_4He2}}
\end{figure*}

\begin{figure*}
	\centering
	\setlength{\unitlength}{0.05\textwidth}
	\begin{picture}(20,8)
	\put(0,0){\includegraphics[clip=true,trim=0.1cm 0cm 0cm 
	0cm,width=0.48\textwidth]{{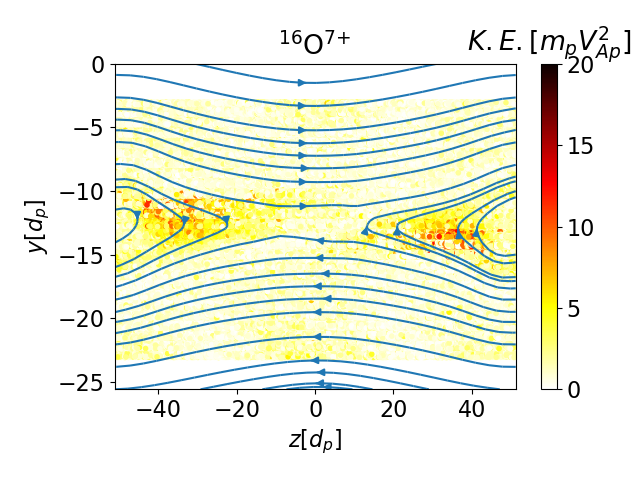}}}
	\put(4,7){{\large $\omega_{cp}t=19.6$}}
	\put(10,0){\includegraphics[clip=true,trim=0.1cm 0cm 0cm 
	0cm,width=0.48\textwidth]{{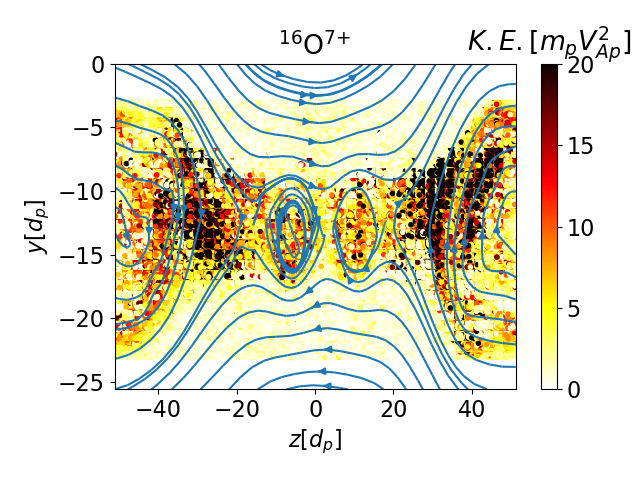}}}
	\put(14,7){{\large $\omega_{cp}t=48.99$}}
	\end{picture}
	\caption{Positions  of ${}^{16}$O$^{7+}$ ions represented by dots colored 
	by ion's kinetic energy at $\omega_{cp}t=19.6$ (left column) and 
	$\omega_{cp}t=48.99$ (right column).  Lines with arrows represent magnetic 
	field lines. Only the lower half ($y < 0$) of the simulation domain is 
	shown. \label{fig:acceleration_sites_16O7}}
\end{figure*}

Heavier ions (${}^{16}$O$^{7+}$ and ${}^{56}$O$^{14+}$ in Fig. 
\ref{fig:av_ke_evolution}) get energized significantly in comparison to the 
lighter Helium ions even during early phase of acceleration ($0 < \omega_{cp}t 
< 30$). This suggests that acceleration mechanisms, different from the 
mechanisms after $\omega_{cp}t=30$, operate in the early phase only on the 
heavier ions, implying a threshold for the energization based on the heaviness 
of the ions.  In order to understand the threshold behavior and the 
acceleration mechanisms before and after $\omega_{cp}t=30$ (the time by which 
$\sim$5\% of the magnetic energy of the asymptotic anti-parallel magnetic field 
has contributed to the development of magnetic islands), we show  the locations 
of the energized particles in the reconnection region at $\omega_{cp}t$=19.6 
and 48.99 in two simulations --- one with  ${}^4$He$^{2+}$ (Fig. 
\ref{fig:acceleration_sites_4He2}) and the other with ${}^{16}$O$^{7+}$ (Fig. 
\ref{fig:acceleration_sites_16O7}). At $\omega_{cp}t=19.6$, ${}^{16}$O$^{7+}$ 
ions are more energized than ${}^4$He$^{2+}$ ions and the locations of the 
energized ${}^{16}$O$^{7+}$ ions are in the outflow regions. The energization 
is, however, not uniformly distributed in the outflow regions: the energized 
particles are located near the upper (lower) separatrix in the left (right) 
outflow region. This observations combined with the threshold behavior of 
energization suggests the pick-up mechanism of the energization in which 
non-adiabatic ions are energized as they cross the reconnection separatrices 
\citep{Drake2009}. A threshold condition,   $m_i/Z_im_p > 5 
\sqrt{2\beta_{p}}/\pi$, for non-adibaticity of ions crossing the sepratrix in 
guide field reconnection was obtained by Drake et al. \citep{Drake2009}, where 
$m_i$ and $Z_i=Q_i/e$ are the mass and charge state of ions and $\beta_p$ is 
the proton plasma beta based on asymptotic value of the anti-parallel magnetic 
field. Although this condition, which becomes $m_i/Z_im_p > 1.12$ for our 
simulation parameters, is satisfied by all the species of heavy ions we have 
considered, the  ${}^4$He$^{2+}$ and ${}^3$He$^{2+}$ ions with $m_i/Z_im_p$ = 2 
and 1.5 were not energized during $\omega_{cp}t<30$.

Note that the threshold condition was obtained assuming that the ions cross the 
separatrix with a velocity equal to $0.1 v_{Ap}$ which is almost the upper 
bound on the inflow velocity in a fully developed steady state magnetic 
reconnection \citep{Liu2017}.  In our simulations, magnetic reconnection at 
$\omega_{cp}t=19.6$ is still developing and the ion inflow velocity has not yet 
reached its maximum value. Using $|u_{iy}| \approx 0.05\,v_{Ap}$, the value of 
inflow velocity at $\omega_{cp}t=19.6$ in our simulations, the threshold 
condition becomes $m_i/Z_im_p > 2.24$ which allows non-adiabatic behavior for 
${}^{56}$Fe$^{14+}$ ($m_i/Z_im_p =4$) and only marginally for ${}^{16}$O$^{7+}$ 
($m_i/Z_im_p = 2.28$) but not for ${}^4$He$^{2+}$ ($m_i/Z_im_p = 2$) and 
${}^3$He$^{2+}$ ($m_i/Z_im_p = 1.5$). 

The localization of energized ${}^{16}$O$^{7+}$ ions near the upper (lower) 
separatrix in the left (right) outflow region at $\omega_{cp}t=19.6$ is due to 
the asymmetries in the lower and upper separatrices in guide field magnetic 
reconnection \citep{Pritchett2004,Li2020}. The thickness of the upper (lower) 
separatrix in the left (right) outflow region is smaller than that of the lower 
(upper) separatrix. This limits the non-adiabaticity and thus energization of 
ions entering outflow regions from the thicker separatrix.

The increased rate of energization for all the heavy ion species after 
$\omega_{cp}t=30$ is somehow linked to the growth of the normal component $B_y$ 
of the magnetic field in the current sheet (Fig. \ref{fig:av_ke_evolution}). 
The normal component $B_y$ in current sheet can grow large by non-steady 
reconnection with increasing rate and/or the compression along the current 
sheet of the reconnected magnetic field lines. The compression can occur  due 
to the pile up of the magnetic flux reconnected at two neighboring sites on the 
magnetic island between the two sites, contraction of magnetic islands and/or 
mutual pushing among magnetic islands. Several acceleration mechanisms may be 
associated with these scenarios: direct acceleration by inductive reconnection 
electric field  in the X-point regions, acceleration by motional electric field 
induced by Alfv\'enic outflow in the outflow regions,  Fermi-like acceleration 
in contracting magnetic islands, magnetic curvature and gradient drifts aligned 
with inductive/motional electric field in the flux pile-up regions and betatron 
acceleration by time dependent magnetic field in the flux compression regions.

Figures \ref{fig:acceleration_sites_4He2} and \ref{fig:acceleration_sites_16O7} 
show that, at $\omega_{cp}t=48.99$, energized ions are located in the X-point 
regions, inside magnetic islands and outflow exhaust regions. Most energized 
particles (black dots in Figures \ref{fig:acceleration_sites_4He2} and 
\ref{fig:acceleration_sites_16O7}) are mostly concentrated near the opening of 
the exhaust regions, where the reconnected magnetic field lines usually pile 
up.  
In the case of simulations with ${}^{16}$O$^{7+}$,
merging of two magnetic islands and presence of energized ions at $z\approx 0$ 
can also be seen  (Fig. \ref{fig:acceleration_sites_16O7}). Although the 
locations of the energetic particles at a given time are not necessarily in the 
neighborhood  of their acceleration sites, it seems from Figs. 
\ref{fig:acceleration_sites_4He2} and \ref{fig:acceleration_sites_16O7}  that a 
number of acceleration mechanisms out of those discussed above are operating 
simultaneously in different regions of reconnection.  A given particle might 
experience acceleration due to different mechanisms  it encounters on its 
trajectory in different reconnection regions. Disentangling these mechanisms 
require a fully history of particles' motion and therefore tracking of their 
trajectories and would be the subject of our future studies.


\begin{figure*}
	\centering
	\setlength{\unitlength}{0.01\textwidth}
	\begin{picture}(100,25)
	\put(0,0){\includegraphics[width=0.32\textwidth]{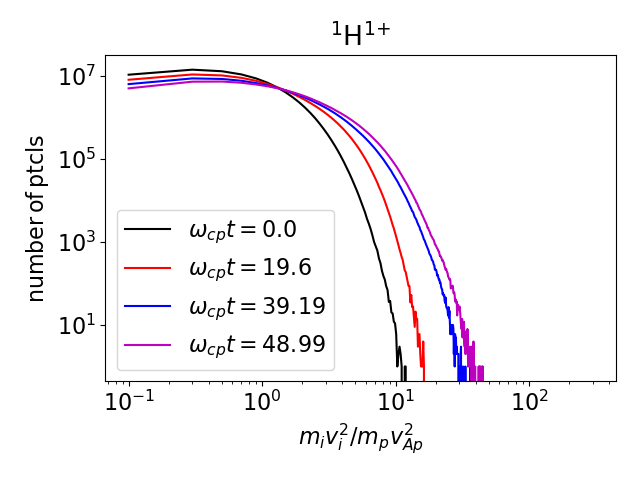}}
	\put(22,18){{\large (a)}}
	\put(33,0){\includegraphics[width=0.32\textwidth]{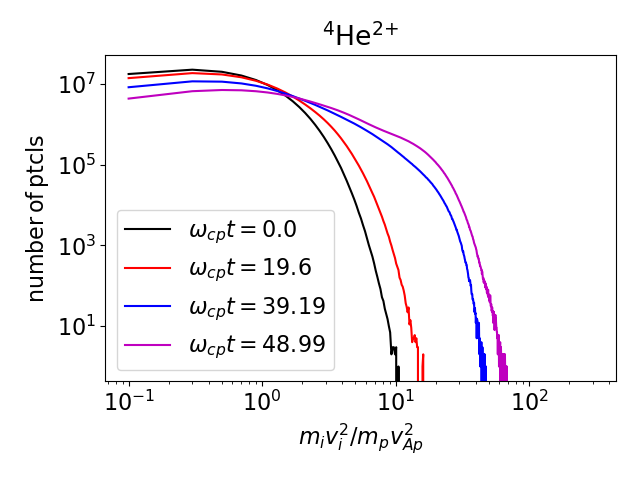}}
	\put(55,18){{\large (b)}}
	\put(66,0){\includegraphics[width=0.32\textwidth]{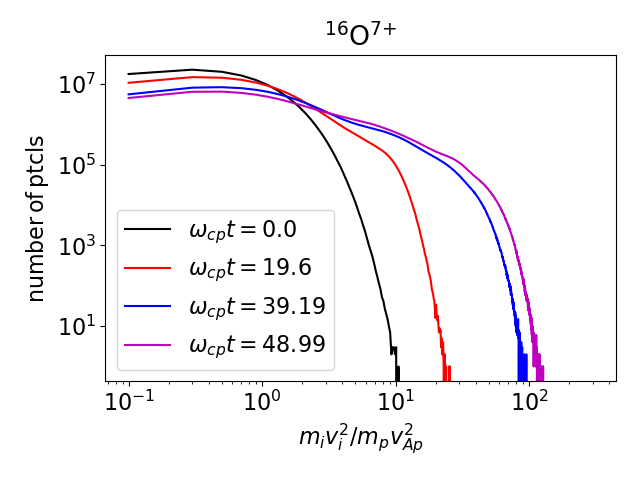}}
	\put(88,18){{\large (c)}}
	\end{picture}
	\caption{Energy spectra of protons ${}^1$H$^{1+}$ (a), ${}^4$He$^{2+}$ (b) 
	and ${}^{16}$O$^{7+}$ (c)  at different times. 
	\label{fig:spectra_evolution}}
\end{figure*}
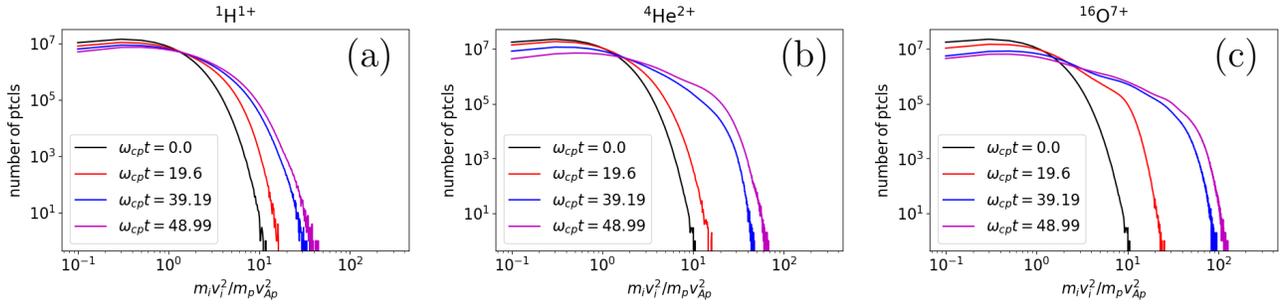

Fig. \ref{fig:spectra_evolution} shows the evolution of the energy spectra for 
different ions species. Energy spectra for all the ion species broaden with 
time.  Energy spectra for protons did not develop any noticeable non-thermal 
feature up to the final simulation time $\omega_{cp}t=48.99$.  Thus energy 
transferred to protons only heat them. Non-thermal feature, a shoulder in the 
spectra in the intermediate energy range after which spectra falls off rapidly, 
develops in the spectra of ${}^{4}$He$^{2+}$ and ${}^{16}$O$^{7+}$ at 
$\omega_{cp}t=39.19$ and $\omega_{cp}t=19.6$, respectively. The early 
development of the spectral shoulder for heavier ions, ${}^{16}$O$^{7+}$ and 
${}^{56}$Fe$^{14+}$ (Figure not shown), in comparison to the lighter ions, 
${}^{4}$He$^{2+}$ and ${}^{4}$He$^{3+}$ (Figure not shown), is consistent with 
the early  energization of heavier ions by pickup mechanism. At 
$\omega_{cp}t=19.6$, the energy of the accelerated  ${}^{16}$O$^{7+}$  (see 
Fig. \ref{fig:acceleration_sites_16O7}) is in the range 1-10 $m_pv_{Ap}^2$ in 
which ${}^{16}$O$^{7+}$ spectra develop a shoulder. The spectral shoulder rises 
with time as well as extends to higher energies  as more and more low energy 
heavy ions are accelerated to higher energies.

\begin{figure*}
	\centering
	\setlength{\unitlength}{0.01\textwidth}
	\begin{picture}(100,35)
	\put(0,0){\includegraphics[width=0.45\textwidth]{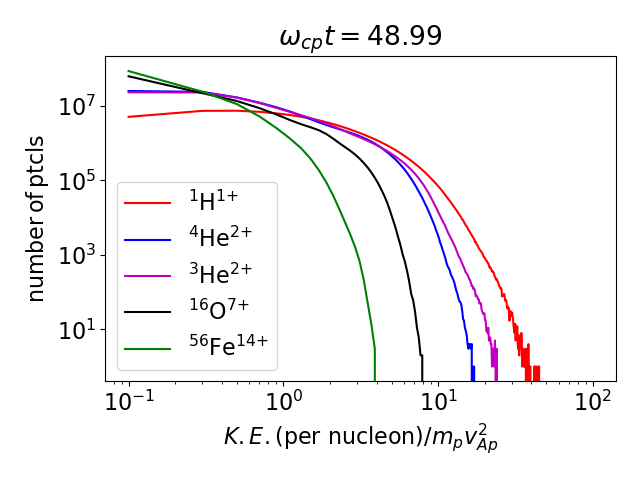}}
	\put(35,25){{\large (a)} }
	\put(50,0){\includegraphics[width=0.45\textwidth]{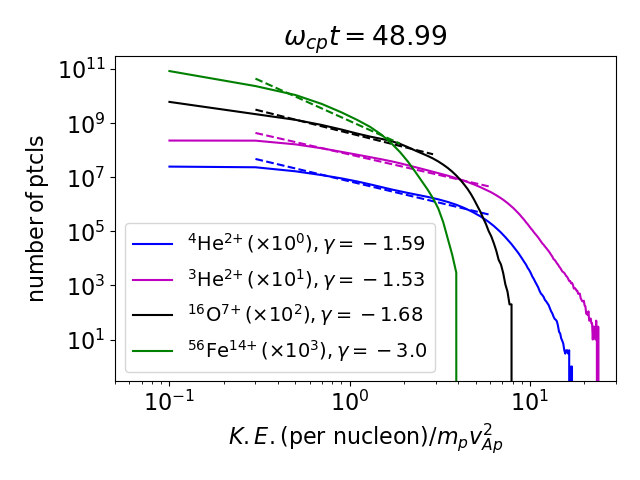}}
	\put(85,25){{\large (b)} }
	\end{picture}
	\caption{Energy-per-nucleon spectra at $\omega_{cp}t=48.99$. Original 
	spectra of different ion species  (a) and spectra of heavy ions shifted for 
	visibility on y-axis by a factor of $10^0$, $10^1$, $10^2$ and $10^3$ for 
	${}^{4}$He$^{2+}$,  ${}^{3}$He$^{2+}$, ${}^{16}$O$^{7+}$ and 
	${}^{56}$Fe$^{14+}$, respectively (b). The dashed lines in the right panel 
	are the power law fit  ($E^{\gamma}$) to the shoulder region of the spectra 
	where $E$ is the energy-per-nucleon. \label{fig:spectra_ions}}
\end{figure*}

\begin{figure*}
	\centering
	\setlength{\unitlength}{0.01\textwidth}
	\begin{picture}(100,35)
	\put(0,0){\includegraphics[width=0.45\textwidth]{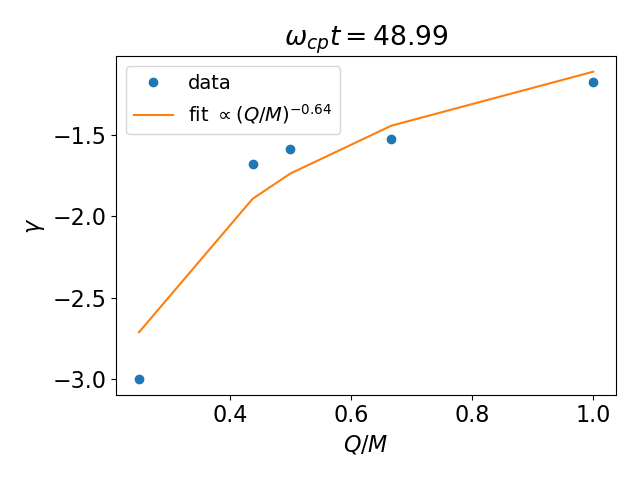}}
	\put(35,8){{\large (a)} }
	\put(50,0){\includegraphics[width=0.45\textwidth]{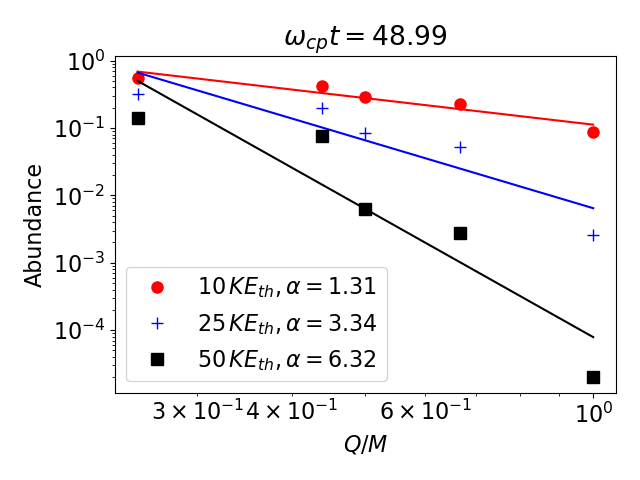}}
	\put(85,8){{\large (b)} }
	\end{picture}
	\caption{ Spectral index $\gamma$ of the energy-per-nucleon spectra in the 
	shoulder region vs. $Q/M$ and the fit $\gamma \propto (Q/M)^{-0.64}$ (a).
		Scaling of the proxy of the ion abundance, defined as number of 
		particles above a threshold energy   normalized to the total number of 
		particles, with $Q/M$ for threshold energy 10 (red filled circles), 25 
		(blue plus signs) and 50 (black filled squares) times the initial 
		thermal energy $KE_{th}$ (b). Straight lines are the fit, abundance 
		$\propto (Q/M)^{-\alpha}$, for the data and are plotted with the color 
		corresponding to the data. \label{fig:qDm_dependence}}
\end{figure*}

Figure \ref{fig:spectra_ions}a compares the energy-per-nucleon spectra of 
different ion species at $\omega_{cp}t=48.99$ --- the last time up to which our 
simulations are valid.
Note that, for the purpose of comparing the spectra of different ion species 
independent of their total mass, the spectra in Fig. \ref{fig:spectra_ions}a is 
shown as a function of energy-per-nucleon unlike the spectra in 
Fig.\ref{fig:spectra_evolution} which is shown as a function of energy. 
The injection energy $\sim 0.3 m_pv_{Ap}^2$, around which the spectral shoulder 
begins, is similar  for heavy ion species. The cutoff energy, up to which the 
shoulder extends, and the maximum gained energy per nucleon are larger for the 
lighter ion species. In Fig. \ref{fig:spectra_ions}b, we plot the same spectra 
(excluding proton's spectra) as in Fig. \ref{fig:spectra_ions}a but 
artificially shifted on y-axis for better visibility of the spectral shoulder 
and power law fits, $E^{\gamma}$, of the shoulder region of the spectra, where 
$E$ is the energy-per-nucleon. The spectral indices of the fits, 
$\gamma$=-1.59, -1.53 and -1.68 respectively for   ${}^{4}$He$^{2+}$,  
${}^{3}$He$^{2+}$ and ${}^{16}$O$^{7+}$ are quite similar. The spectra for 
${}^{56}$Fe$^{14+}$, on the other hand, is somewhat softer with $\gamma=-3.0$. 
The dependence of $\gamma$ on $Q/M$, shown in Fig. \ref{fig:qDm_dependence}a, 
can be approximately fitted as $\gamma \propto (Q/M)^{-0.64}$. 


In ${}^{3}$He-rich SEP events, ion abundances are enhanced relative to their 
solar abundances at energies well above their thermal energy. We consider a 
proxy for abundance enhancement factor $F$ as number of particles above a 
threshold energy normalized to the total number of particles. Fig. 
\ref{fig:qDm_dependence}b shows this proxy of the abundance enhancement factor 
as a function of $Q/M$ at $\omega_{cp}t=48.99$ for three energy thresholds 
$E_t=10 \, \mathit{KE}_{th}, 25 \, \mathit{KE}_{th}$ and $50\, 
\mathit{KE}_{th}$ well above the initial thermal energy $\mathit{KE}_{th}=0.25 
m_pv_{Ap}^2$ which is the same for all ion species in our simulations. The 
value of $E_t$ is so chosen that it is  in the tail region of the initial 
energy spectra which is the same for all the ion species and at the same time 
is not outside the shoulder regions of the energy spectra of the heavy ions at 
$\omega_{cp}t=48.99$. Although the value $E_t=50\, \mathit{KE}_{th}=12.5 
\,m_pv_{Ap}^2$ is close to the cutoff energy of the shoulder region in the 
energy spectra of ${}^4$He$^{2+}$ at $\omega_{cp}t=48.99$ (see Fig. 
\ref{fig:spectra_evolution}b), the abundances will be dominated by the 
abundances in the shoulder region as the energy spectra falls off very rapidly 
beyond the shoulder region. Note that the threshold $E_t=12.5 \,m_pv_{Ap}^2$ is 
in the rapidly falling region of the proton's energy spectra at 
$\omega_{cp}t=48.99$. It is, however, acceptable as proton's energy spectra 
does not develop a shoulder.             

Fig. \ref{fig:qDm_dependence}b also shows the power law fit $F \propto 
(Q/M)^{-\alpha}$. The enhancement factor drops with increasing $Q/M$, i.e., 
heavier ions are preferentially accelerated. The drop of the enhancement factor 
becomes steeper (the value of $\alpha$ increases) with the increasing threshold 
energy. 
As the value of $E_t$ increases, the contribution of the shoulder region 
relative to the contribution of the rapidly falling region of the energy 
spectra towards the abundances decreases. This decrease in the contribution  is 
more prominent for larger values of $Q/M$ as the shoulder in their energy 
spectra extends to smaller energies. This results in the increase in the value 
of $\alpha$ with $E_t$.

\section{Discussion and conclusion  \label{sec:conclusion}}
We carried out hybrid-kinetic simulations (with electron inertia) to study the 
acceleration of heavy ions ($Q/M < 1$) by magnetic reconnection. We find that 
heavy ions can be accelerated to high energies many times larger than their 
initial energies by a variety of acceleration mechanisms. Heavier ions are 
preferentially accelerated in the sense that energy gain averaged over 
particles increases with decreasing $Q/M$.
They are primarily accelerated in magnetic islands and flux pile up regions 
near the opening of the outflow exhausts. Most efficient acceleration takes 
place in the flux pileup regions. Heavy ions, depending upon the smallness of 
$Q/M$ which allows them to be non-adiabatic while crossing from inflow to 
outflow regions, can also be accelerated by pickup mechanism in outflow regions 
even before any magnetic flux is piled up. As a result of acceleration, heavy 
ions develop a shoulder, a non-thermal feature, in their energy spectra. The 
spectral index obtained from the power law fit in the shoulder region of the 
spectra varies approximately as $(Q/M)^{-0.64}$. Abundance enhancement factor, 
defined as number of particles above a threshold energy normalized to total 
number of particles, scales as $(Q/M)^{-\alpha}$ where $\alpha$ increases with 
the energy threshold. 


Energy spectra with a shoulder or in other words double power law with a break 
in the spectra  at  energy $\sim$ 1 MeV/nucleon has been in-situ observed  in 
space \citep{Mason2002,Bucik2018}. Our simulations show the break in the energy 
spectra in the energy range 2-6 $m_pv_{Ap}^2$ per nucleon depending upon the 
ion species.  The value of $v_{Ap}$ in the active regions of solar corona, 
where acceleration takes place, has been estimated to lie in the range 
2000-9000 km/s \citep{Brooks2021}. For a typical value $v_{Ap}$=5000 km/s, 
$m_pv_{Ap}^2\approx $0.25 MeV and therefore break in the simulations occurs in 
the range 0.5-1.5 MeV/nucleon, consistent with the observations. Note that the 
energy range of the break depends on the value of $v_{Ap}$, estimates of which 
in solar corona vary significantly \citep{Brooks2021}. In the observations, the 
spectral index $\gamma$ of the power law before the break is in the range 1-3 
for the ions of Helium-4, Helium-3, Oxygen and Iron. The values of $\gamma$ 
before the break is in the same range in our simulations as well. The values of 
$\gamma$ after the break are, however, much larger in our simulation in 
comparison to the observations.  

We defined  a proxy $F$ for the abundance enhancement factor to study its 
variation with $Q/M$. Although this proxy does not exactly correspond to the 
abundance enhancement factor used in observations, we point out some similarity 
and differences in the behavior of the two. In our simulations, the power law 
index $\alpha$ of the fit $F\propto (Q/M)^{-\alpha}$  increases with the 
threshold energy. Observation also show event to event variation in the value 
of $\alpha$ in the range $-5 < \alpha < 12$ with a mean value of $\alpha=$3.26 
at 385 keV/nucleon \citep{Mason2004} and $2< \alpha < 8$ with a mean value of 
$\alpha=3.64$ at 3-5 MeV/nucleon \citep{Reames2014a,Reames2014}. The mean 
values in the observations are similar despite their energies being an order of 
magnitude apart. The mean of the three values of $\alpha$ obtained from power 
law fits in Fig. \ref{fig:qDm_dependence}b is approximately 3.66. In 
observations, steeper energy spectra tend to have steeper fall of the abundance 
enhancement with $Q/M$, i.e., large value of $\alpha$ \citep{Reames2014a}. This 
effect is similar to the increase in the value of $\alpha$ with $E_t$ in our 
simulations. Larger values of $E_t$ increases the contribution of the steeper 
part of the energy spectra in the calculations of the abundances.


 Our simulations consider only a single species of heavy ions at a time. In a physical situation, multiple heavy ion species will be present in  reconnection region at a same time and accelerated simultaneously by magnetic reconnection, partitioning the available magnetic energy among the species. This can affect the maximum energy gained by a given species but is unlikely to affect the preferential acceleration of heavy ions, at least as long as the heavy ions are minor species ( in the solar corona, heavy ions with $Z>2$  compose only 0.1\% of all ions while Helium is 8.9\% \citep{lodders2020}) so that the electric and magnetic field in the reconnection region are mainly due to the electron and protons dynamics with only a small contribution from the heavy ion species. Simulations of magnetic reconnection with multiple heavy ion species are, however, required to understand if the minority species can influence the accelerating fields and, if yes, to what extent.    

Note that the SEP events are detected in space far away from their acceleration 
sites on Sun. The observed scaling of abundance enhancement with $Q/M$ and 
other features of these events may, therefore, not necessarily be the effect of 
only acceleration but also of the transport of particles from the acceleration 
site to the detection site. In fact, 3-D test particle modeling of the 
inter-planetary transport of relativistic protons from the source region on Sun 
shows that the spectra of the particles is highly observer dependent and do not 
necessarily reflect the source spectra \citep{Dalla2020}. Such transport 
effects are possible for the spectra of heavy ions as well. 
	However, observations of impulsive events (their intensity time profiles, 
	velocity dispersion, and anisotropies) indicate their nearly scatter-free 
	propagation with scattering mean free paths $> 1$ au (e.g., 
	\cite{reames2020, reames2022} and therefore a direct comparison with 
	acceleration models is 
	meaningful (\cite{Mason2002} and references therein). The authors point 
	out that while the effects of interplanetary propagation on these events 
	are not large, definitive resolution requires observations close to the 
	Sun.

We conclude that magnetic reconnection is a potential candidate for the 
	preferential acceleration of heavy ions which may provide  a power law 
	dependence of the abundance enhancement on $Q/M$, as suggested by the 
	results 
	presented here. More detailed studies on the relative roles of the 
	different 
acceleration mechanisms operating in magnetic reconnection in the abundance 
enhancement will be presented in a future publication. Note that we did not 
find any extra-ordinary abundance enhancement of ${}^3$He$^{2+}$ in our 
simulations. This, however, does not rule out the role of magnetic reconnection 
in the abundance enhancement of ${}^3$He$^{2+}$ as our simulations are carried 
out only for limited range of parameters and  in two dimensions. Magnetic reconnection in three dimensions is a host to a number of additional three-dimensional instabilities and allows more complex interactions in the reconnection region. For example, flux ropes formed by the growth of a 3-D oblique plasmoid instability in a low-beta reconnecting current layer undergo complex interactions due to the advection  and rotation by the reconnection outflow jets, leading to a turbulent state with stochastic magnetic field and development of secondary flux ropes in thin current layers that form between the primary flux-ropes \citep{stanier2019}. A finite guide magnetic field of the order of the reconnecting component and axial structure of flux ropes in 3-D magnetic reconnection enable efficient electron acceleration by allowing the particles to leak out and return to acceleration regions, as shown by 3-D kinetic simulations \citep{dahlin2017}.
Fully kinetic simulations show that electron acceleration  by parallel electric field and curvature driven mechanisms can be enhanced by the filamentation of electric and magnetic fields caused by streaming instabilities in the nonlinear stage of 3-D guide-field magnetic reconnection \citep{munoz2018b}. 
Strong drift-wave fluctuations in the lower-hybrid frequency range, developed in fully kinetic 3-D simulations of asymmetric magnetic reconnection (modeling MMS magnetopause diffusion region crossings), give rise to  cross-field electron transport and significantly enhanced electron parallel heating in comparison to 2-D simulations \citep{le2018}. It seems likely that 3-D plasma instabilities in magnetic reconnection will interact with  ${}^3$He$^{2+}$ and other heavy ions as well and contribute to their  acceleration, as they do for electron acceleration.  The details of these complex interactions are unknown and subject for further studies. 


\begin{acknowledgements}
	The authors thank Patricio Mu\~noz for fruitful discussion and his help. We 
	gratefully acknowledge  the financial support by the German Science 
	Foundation (DFG), projects JA 2680-2-1 and BU 777-17-1 as well as the Czech 
	Republic  GA\^CR project  20-09922J M, personally Markus Rampp and Meisam 
	Tabriz  of the Max Planck Computing and Data Facility (MPCDF) for their 
	support of  the development  of the hybrid-kinetic code CHIEF used for the 
	simulation studies  presented in this paper. The simulations were carried 
	out on the MPS supercomputers at the MPCDF, Garching,  Germany. R.B. 
	acknowledges support by NASA grants 80NSSC21K1316 and
	379 80NSSC22K0757.	
\end{acknowledgements}

%
   \bibliographystyle{aa} 
   \bibliography{references_HIA} 
%

\end{document}